\documentclass[english,aps,reprint]{revtex4-1}
\usepackage[T1]{fontenc}
\usepackage[latin9]{inputenc}
\usepackage{geometry}
\usepackage{xcolor}
\usepackage{babel}
\usepackage{array}
\usepackage{longtable}
\usepackage{booktabs}
\usepackage{amssymb}
\usepackage{graphicx}
\usepackage{esint}
\PassOptionsToPackage{normalem}{ulem}
\usepackage{ulem}
\usepackage[unicode=true,
 bookmarks=false,
 breaklinks=false,pdfborder={0 0 1},backref=section,colorlinks=false]
 {hyperref}

\makeatletter

\providecommand{\tabularnewline}{\\}
\providecolor{lyxadded}{rgb}{0,0,1}
\providecolor{lyxdeleted}{rgb}{1,0,0}
\DeclareRobustCommand{\lyxsout}[1]{\ifx\\#1\else\sout{#1}\fi}

\usepackage{babel}

\makeatother

\begin{document}
\title{Anomalous Temperature Dependence of Quantum Correction to the Conductivity
of Magnetic Topological Insulators}
\author{Huan-Wen Wang}
\affiliation{Department of Physics, The University of Hong Kong, Pokfulam Road,
Hong Kong, China}
\author{Bo Fu}
\affiliation{Department of Physics, The University of Hong Kong, Pokfulam Road,
Hong Kong, China}
\author{Shun-Qing Shen}
\email{sshen@hku.hk}

\affiliation{Department of Physics, The University of Hong Kong, Pokfulam Road,
Hong Kong, China}
\begin{abstract}
Quantum transport in magnetic topological insulators reveals strong
interplay between magnetism and topology of electronic band structures.
A recent experiment on magnetically doped topological insulator $\mathrm{Bi_{2}\mathrm{Se_{3}}}$
thin films showed the anomalous temperature dependence of the magnetoconductivity
while their field dependence presents a clear signature of weak anti-localization
{[}Tkac \emph{et al}., PRL 123, 036406(2019){]}. Here we demonstrate
that the tiny mass of the surface electrons induced by the bulk magnetization
leads to a temperature-dependent correction to the $\pi$ Berry phase,
and generates a decoherence mechanism to the phase coherence length
of the surface electrons. As a consequence, the quantum correction
to conductivity can exhibit non-monotonic behavior by decreasing temperature.
This effect is attributed to the close relation of the Berry phase
and quantum interference of the topological surface electrons in quantum
topological materials.
\end{abstract}
\maketitle

\paragraph*{Introduction.}

\begin{figure}
\centering{}\includegraphics[clip,width=8cm]{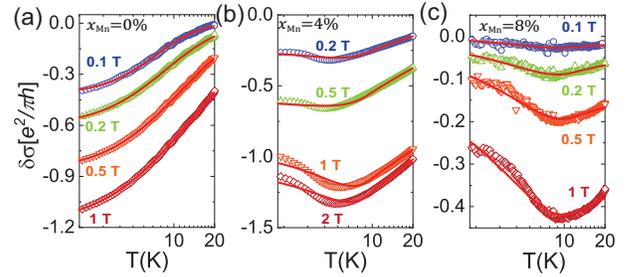}\caption{\label{fig:anomalous conductivity}Magnetoconductivity as a function
of temperature at different magnetic field strength for two Mn-doped
$\mathrm{Bi_{2}Se_{3}}$ thin films of Mn-doped concentration (a)
$x_{\mathrm{Mn}}=0\%$, (b) $x_{\mathrm{Mn}}=4\%$ , and (c) $x_{\mathrm{Mn}}=8\%$.
The open squares are the experimental data extract from Ref. \citep{tkavc2019influence}.
The solid red lines are the fitting results at different magnetic
filed $B$ by using the formula in Eq. (\ref{eq: MC_single band}).}
\end{figure}

Three-dimensional (3D) topological insulators (TIs) have stimulated
intensive theoretical and experimental study in the past decade \citep{hasan2010colloquium,qi2011topological,moore2010birth,ando2013topological,SQS,culcer2019transport}.
In the quantum diffusive regime, owing to the nontrivial $\pi$ Berry's
phase, the topological surface states are expected to experience a
destructive quantum interference in the scattering process \citep{ando1998berry,suzuura2002crossover,Lu2011WAL,lu2014weak}.
Accordingly, the magnetoconductivity shows a negative notch in a weak
magnetic field ($B$) and is called weak anti-localization (WAL),
which has been regarded as significant transport signature for the
topological surface states of TIs \citep{checkelsky2009quantum,chen2010gate,HeHT2011prl,steinberg2011electrically,PhysRevB.83.241304,kim2011thickness}.
Besides, one anticipates that the conductivity correction from the
WAL effect should decrease with increasing the temperature. However,
the temperature dependence of conductivity usually shows an opposite
tendency in experiments \citep{Liu2011prb,WangJ2011prb,takagaki2012weak,liu2014tunable,wang2016thickness,jing2016weak}.
Such a dilemma in some pristine TIs can be resolved by further considering
the electron-electron interaction effect at low temperatures \citep{Altshuler1980interaction,PatrickLee1985rmp,Lu_finite_T_prl_2014}.
Recently, Tkac \emph{et al}. reported that the contradictory tendency
between the temperature- and magnetic-field-dependent conductivity
remains even after subtracting the interaction effect in the Mn-doped
$\mathrm{Bi_{2}Se_{3}}$ thin films \citep{tkavc2019influence}. As
shown in Fig. \ref{fig:anomalous conductivity}, the magnetoconductivity
$\delta\sigma(B)$ exhibits monotonic temperature dependence for a
non-doped $\mathrm{Bi_{2}Se_{3}}$ sample, a typical behavior of WAL
as expected theoretically, and a non-monotonic temperature dependence
for the doped ($x_{\mathrm{Mn}}=4\%$ and $x_{\mathrm{Mn}}=8\%$)
samples, respectively, where $\delta\sigma(B)=\sigma(T,B)-\sigma(T,0)$
with $\sigma(T,B)$ the temperature-dependent conductivity at a finite
magnetic field $B$. At low temperatures, the doped and non-doped
samples show opposite temperature dependence. Meanwhile, the magnetoconductivity
for those samples always exhibit WAL correction as shown in Fig. 2
in \citep{tkavc2019influence}. The simple assumption of the monotonic
temperature dependence of coherence length due to the electron-electron
interaction effect \citep{Altshuler1980interaction,PatrickLee1985rmp,efros1985electron}
cannot account for these observations. Actually, the surface state
in the magnetically doped TIs acquires a finite mass due to the time-reversal
symmetry breaking accompanied with a small correction to the $\pi$
Berry phase \citep{LiuQ2009massive,PhysRevB.81.195203,chen2010massive,Liu_crossover_prl_2011,tokura2019magnetic}.
The nearly $\pi$ Berry phase is capable of accounting for the WAL
behavior for the magnetoconductivity but fails to explain the anomalous
behavior.

In this Letter, we resolve the puzzle of the anomalous temperature
dependence of quantum correction. The role of the magnetic doping
is assumed to produce a finite gap for the surface states. Then, a
magnetoconductivity formula of quantum interference is derived for
massive Dirac fermions, which is simply characterized by the spin
polarization $\eta$. The quantity is also associated to the correction
to the $\pi$ Berry phase of surface electrons. The nearly $\pi$
Berry phase accounts for the WAL behavior for the magnetoconductivity.
However, the temperature dependence of $\eta$ leads to a non-monotonic
behavior of the quantum correction to the conductivity at low temperatures
due to the quantum decoherence effect caused by the deviation from
the $\pi$ Berry phase. The good coincidence between our theory and
experimental data suggests that the anomalous temperature dependence
can be ascribed to the temperature-dependent correction to the $\pi$
Berry phase of the surface states.

\paragraph*{Model Hamiltonian and spin polarization.}

\begin{figure}
\begin{centering}
\includegraphics[width=8cm]{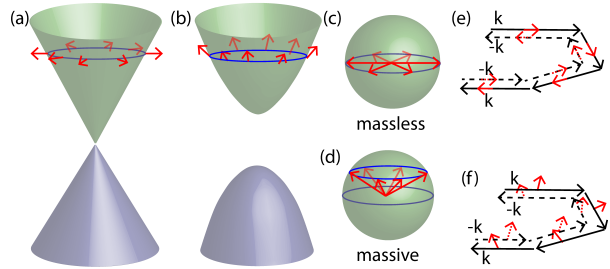} 
\par\end{centering}
\caption{\label{fig:berry phase} Schematic diagram of the band structure and
spin orientation for (a) massless and (b) massive Dirac fermions.
The spin vectors at a certain Fermi energy are depicted by the red
arrows. (c) and (d) show the corresponding Berry phase as the solid
angle traced out the spin vectors on the Bloch sphere for (a) and
(b), respectively. (e) and (f) show the trajectory of backscattering
(solid line) and corresponding time-reversal trajectory (dashed line)
for massless and massive Dirac fermions, respectively. The black arrow
represent the momentum direction, and the red arrow denotes the spin
orientation.}
\end{figure}

Due to the hybridization of the top and bottom surface states or the
time-reversal symmetry breaking caused by the magnetic doping, the
surface electrons in the TI thin films can acquire a finite mass \citep{LiuQ2009massive,shan2010effective,lu2010massive,zhang2010crossover},
thus it is proper to treat the surface states as massive Dirac fermions.
Besides, in a TI thin film, the 3D bulk band is quantized into two-dimensional
(2D) sub-bands owing to the quantum confinement effect. The 2D sub-bands
have a similar low energy Hamiltonian as the surface one but with
a relatively large band gap \citep{Lu2011bulk}. We begin with the
modified model of 2D massive Dirac fermions \citep{SQS,lu2010massive},
\begin{equation}
H=v\hbar(\sigma_{x}k_{x}+\sigma_{y}k_{y})+m(k)\sigma_{z}\label{eq:2dmassive}
\end{equation}
where $v$ is the effective velocity, $\hbar$ is the reduced Planck
constant, $\sigma_{x,y,z}$ are the Pauli matrices, $\mathbf{k}=(k_{x},k_{y})$
is the wave vector, and $m(k)=mv^{2}-b\hbar^{2}(k_{x}^{2}+k_{y}^{2})$
is the mass term, and $m$ and $b$ are the coefficients. The mass
term gives the spin polarization $\eta=\langle\sigma_{z}\rangle=m(k_{F})/\sqrt{v^{2}\hbar^{2}k_{F}^{2}+[m(k_{F})]^{2}}$
at the Fermi radii $k_{F}$, which is directly related to the Berry
phase for Dirac fermions. As shown in Fig. \ref{fig:berry phase},
the spin lies in the plane of the Fermi circle for $\eta=0$ and is
titled to the out of plane for $\eta\ne0$. After the spin vector
travels along the Fermi circle adiabatically, a Berry phase is acquired,
$\phi_{b}=\frac{1}{2}\int_{0}^{\pi}\int_{0}^{\arccos\eta}\sin\theta d\theta d\phi=\pi(1-\eta)$.
Furthermore, we mark the spin and momentum orientation in the trajectory
of backscattering and corresponding time-reversal trajectory. For
$\eta=0$, the spins of incoming ($\mathbf{k}$) and outgoing ($-\mathbf{k}$)
electrons are anti-parallel to each other. The scattering sequences
are accompanied by the coherent spin rotation which yields the WAL
due to the $\pi$ Berry phase. For $\eta\ne0$, the spin of the $(\mathbf{k},-\mathbf{k}$)
electron pair is partially titled to the $z-$direction, and the spin-singlet
and triplet pairings mix together. Consequently, the accumulating
Berry phase deviates from $\pi$, and after taking the average of
all the possible trajectories with different winding numbers, a new
decoherence mechanism is introduced. When $\eta\to1$, the spin is
along the $z$-direction. The incoming and outgoing electrons form
a triplet pairing and give rise to a WL correction.

\paragraph*{Cooperon gaps and weighting factors.}

\begin{table}
\centering{}\caption{\label{tab:The-components-of}The components of four Cooperon channels
$i=s,t_{0,\pm}$ in the basis of spin-triplet and singlet $|s,s_{z}\rangle$,
the Cooperon gap $\ell_{i}^{-2}$ in unit of the mean free path $\ell_{e}^{-2}$
and the weighting factors $w_{i}$.}
\begin{longtable}[c]{c>{\centering}p{3cm}>{\centering}p{2cm}>{\centering}p{2cm}}
\toprule 
$i$  & Cooperon in $|s,s_{z}\rangle$  & $w_{i}$  & $\ell_{i}^{-2}/\ell_{e}^{-2}$\tabularnewline
\endhead
\midrule 
\addlinespace
$s$  & $|0,0\rangle$  & $-\frac{(1-\eta^{2})^{2}}{2(1+3\eta^{2})^{2}}$  & $\frac{(1-\eta^{2})\eta^{2}}{(1+\eta^{2})^{2}}$\tabularnewline
\addlinespace
$t_{+}$  & $|1,1\rangle$  & $\frac{4\eta^{2}(1+\eta^{2})}{(1+3\eta^{2})^{2}}$  & $\frac{4(1-\eta)^{2}\eta^{2}}{(1+3\eta^{2})(1+\eta)^{2}}$\tabularnewline
\addlinespace
$t_{0}$  & $|1,0\rangle$  & $0$  & $\infty$\tabularnewline
\addlinespace
$t_{-}$  & $|1,-1\rangle$  & $\frac{4\eta^{2}(1+\eta^{2})}{(1+3\eta^{2})^{2}}$  & $\frac{4(1+\eta)^{2}\eta^{2}}{(1+3\eta^{2})(1-\eta)^{2}}$\tabularnewline
\bottomrule
\end{longtable}
\end{table}

The quantum correction to the conductivity is evaluated by using the
Feynman diagrammatic technique \citep{gor1979particle,hikami1980spin,bergmann1984weak,McCann2006weaklocalization,Garate2012bulksurface,fu2019_prl_quantum}.
In the present calculation, we keep the matrix form for Green's functions
and treat all possible Cooperon channels, correlators in the particle-particle
pairing channels in electric conductivity of non-superconducting metals,
on the same footing \citep{Note-on-SM}. In the diffusion approximation,
it is found that three out of four possible Cooperon channels contribute
to the conductivity, 
\begin{equation}
\sigma_{\mathrm{qi}}=-\frac{4e^{2}}{h}\sum_{i}\sum_{q}\frac{w_{i}}{\ell_{i}^{-2}+q^{2}}\label{eq:sigmaqi}
\end{equation}
where $i=s,t_{+},t_{-}$ is the Cooperon channel index, $\ell_{i}^{-2}$
and $w_{i}$ are the corresponding Cooperon gap and weighting factors,
respectively. The expressions for $\ell_{i}^{-2}$ and $w_{i}$ are
listed in Table \ref{tab:The-components-of}.

\begin{figure}
\begin{centering}
\includegraphics[clip,width=8cm]{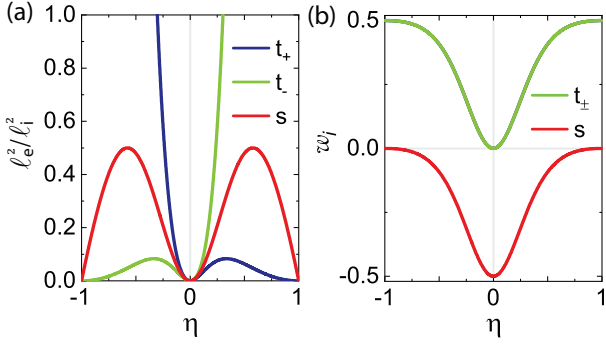} 
\par\end{centering}
\caption{\label{cooperon gap and weighting factors}(a) The Cooperon gap $\ell_{i}^{-2}$
in the unit of square of the mean free path $\ell_{e}^{-2}$ and (b)
the weighting factors as functions of spin polarization $\eta$, where
$t_{0,\pm}$ and $s$ represent the WL and WAL channels, respectively.
The weighting factors for $t_{+}$ and $t_{-}$ channels are equal.}
\end{figure}

The channels $i=t_{\pm}$ contribute to the WL correction, and the
channel $s$ contributes to the WAL correction according to the signs
of their weighting factors $w_{t_{\pm}}>0$ and $w_{s}<0$. The original
Cooperon structure factor $\Gamma(q)$ is in the basis of $\{\left|\uparrow\uparrow\right\rangle ,\left|\uparrow\downarrow\right\rangle ,\left|\downarrow\uparrow\right\rangle ,\left|\downarrow\downarrow\right\rangle \}$.
To diagonalize $\Gamma(q)$, we rotated the basis into the spin-singlet
and triplet basis $|s,s_{z}\rangle$, where $|s,s_{z}\rangle$ labels
the total spin $s(=0,1)$ and its $z$-component $s_{z}$. The channels
$i=t_{\pm}$ correspond to the two triplet pairing ($s=1$) and result
in the WL correction, while the channel $i=s$ is the singlet pairing
($s=0$) and gives out the WAL correction. $\ell_{i}^{-2}$ and $w_{i}$
are plotted in Fig. \ref{cooperon gap and weighting factors}. When
$\eta=0$ and $\phi_{b}=\pi$, one finds a pure WAL correction from
the channel $s$, which is consistent with the Hikami-Larkin-Nagaoka
formula for the strong spin-orbit scattering \citep{hikami1980spin}.
When $\eta=1$ $(\eta=-1)$ and $\phi_{b}=0$ ($2\pi$), the channel
$t_{+}$($t_{-}$) gives a pure WL correction as the conventional
electron gas.

\paragraph*{Temperature dependence of conductivity correction.}

\begin{figure}
\begin{centering}
\includegraphics[clip,width=8cm]{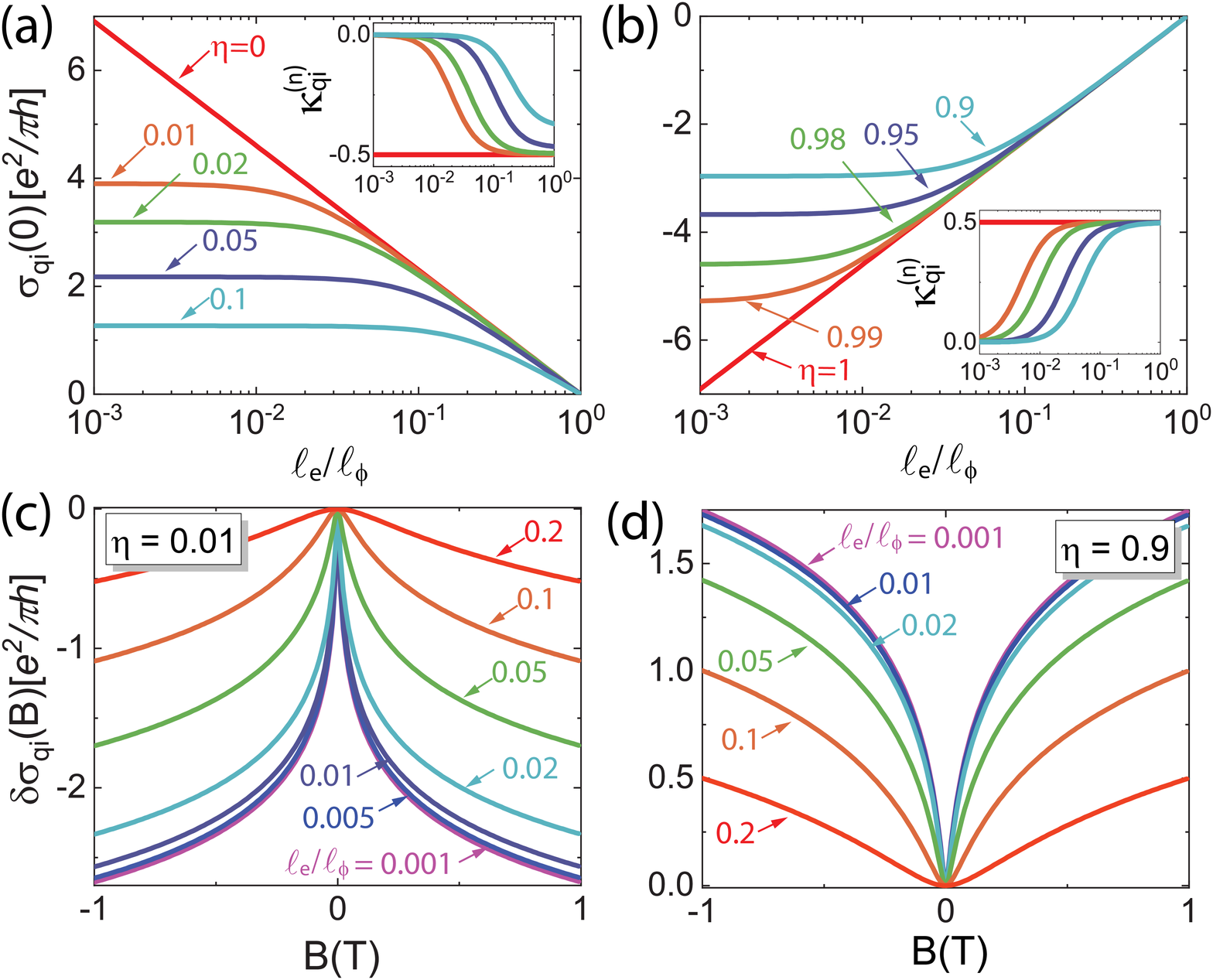} 
\par\end{centering}
\caption{\label{fig:zero-field-conductivity-correcti}Zero-field conductivity
correction and slope $\kappa_{\mathrm{qi}}^{(n)}$ as a function of
the ratio of the mean free path to the coherence length $\ell_{e}/\ell_{\phi}$
for (a) WAL of spin polarization $\eta\sim0$ and (b) WL of $\eta\sim1$.
Magnetoconductivity at different values of $\ell_{e}/\ell_{\phi}$
for (c) $\eta=0.01$ and (d) $\eta=0.9$. The calculation parameter
$\ell_{e}=10\mathrm{nm}$.}
\end{figure}

The integration over $q$ in Eq. (\ref{eq:sigmaqi}) is logarithmically
divergent in both the ultraviolet and ultra-infrared limit. To avoid
the divergence, the two cut-offs have to be introduced to restrict
$\ell_{\phi}^{-1}\leq q\le\ell_{e}^{-1}$, where $\ell_{e}=\sqrt{\mathcal{D}_{0}\tau}$
is the mean free path and $\ell_{\phi}$ is the coherence length caused
by the inelastic scattering \citep{Altshuler1980interaction,PatrickLee1985rmp,efros1985electron}.
Consequently, Eq. (\ref{eq:sigmaqi}) gives the quantum correction
to the conductivity, 
\begin{equation}
\sigma_{\mathrm{qi}}(B=0,T)=\frac{e^{2}}{\pi h}\sum_{i}w_{i}\ln\frac{\ell_{\phi}^{-2}+\ell_{i}^{-2}}{\ell_{e}^{-2}+\ell_{i}^{-2}}.
\end{equation}
To investigate the temperature dependence of $\sigma_{qi}(T)$, we
assume $\ell_{\phi}=\ell_{\phi}^{0}(T/T_{0})^{-p/2}$, where $p=1$
for electron-electron interaction and $p=3$ for electron-phonon interaction
in 2D systems, $\ell_{\phi}^{0}$ is the coherence length at $T=T_{0}$
\citep{Altshuler1980interaction,PatrickLee1985rmp}. The characteristic
parameter of the temperature-dependent conductivity is \citep{Lu_finite_T_prl_2014}
\begin{equation}
\kappa_{\mathrm{qi}}^{(n)}\equiv\frac{\pi h}{e^{2}}\frac{\partial\sigma_{\mathrm{qi}}(B=0,T)}{\partial\ln T}=\sum_{i}\frac{w_{i}p}{1+\ell_{\phi}^{2}/\ell_{i}^{2}}\label{kappa-2}
\end{equation}
if $\eta$ and $\ell_{e}$ are insensitive to the temperature. In
this case, the presence of non-zero Cooperon gap $\ell_{i}^{-2}$
is highly non-trivial. As shown in Fig. \ref{fig:zero-field-conductivity-correcti}(a),
when $\eta=0$, the conductivity correction is always logarithmically
divergent and $\kappa_{\mathrm{qi}}^{(n)}=-p/2$. However, once $0<\eta\ll1,\ell_{i}^{-2}\ne0$,
the conductivity correction saturates at lower temperatures and $\kappa_{\mathrm{qi}}^{(n)}$
would increase from some value $\in(-p/2,0)$ to $0$ gradually. In
another limit of $\eta\sim1$, as shown in Fig. \ref{fig:zero-field-conductivity-correcti}(b),
$\kappa_{\mathrm{qi}}^{(n)}=p/2$ for $\eta=1$, and $\kappa_{\mathrm{qi}}^{(n)}$
decreases from some value $\in(0,p/2)$ to $0$ by lowering temperature.
Hence, the finite Cooperon gap leads to the saturation behavior of
$\sigma_{\mathrm{qi}}(0)$ at low temperatures.

In the magnetic TIs, the mass term is related to the magnetization,
hence $\eta$ is also a function of temperature. Consequently, the
slope $\kappa_{\mathrm{qi}}$ has a correction term from $\partial\eta/\partial\ln T$,
\begin{equation}
\kappa_{\mathrm{qi}}^{(m)}=\sum_{i}\left(g_{i}\frac{\partial\eta}{\partial\ln T}+\frac{w_{i}p}{1+\ell_{\phi}^{2}/\ell_{i}^{2}}\right)\label{kappa}
\end{equation}
with $g_{i}\equiv\frac{\partial}{\partial\eta}(w_{i}\ln\frac{\ell_{\phi}^{-2}+\ell_{i}^{-2}}{\ell_{e}^{-2}+\ell_{i}^{-2}})$.
Here we still assume that $\ell_{e}$ is insensitive to the temperature.
We can have a qualitative analysis for the sign of $\kappa_{qi}$
for the case of $\eta\sim0$. When $\eta\sim0$, $\kappa_{\mathrm{qi}}^{(m)}\approx-\frac{\ell_{e}^{-2}\partial\eta^{2}/\partial\ln T+\ell_{\phi}^{-2}p}{2(\ell_{\phi}^{-2}+\ell_{i=s}^{-2})}$.
If $\frac{\partial\eta}{\partial\ln T}\ge0$ and $\kappa_{\mathrm{qi}}\le0$,
the zero-field conductivity always decreases with increasing temperature,
indicating a WAL tendency as usual. However, if $\frac{\partial\eta}{\partial\ln T}<0$,
$\ell_{e}^{-2}\frac{\partial\eta^{2}}{\partial\ln T}<0$ and $\ell_{\phi}^{-2}p>0$.
$\kappa_{\mathrm{qi}}$ may experience a sign change while decreasing
temperature, which implies anomalous temperature dependence even in
the case of the WAL correction. A similar analysis holds for $\eta\sim1$.

\paragraph*{Magnetoconductivity.}

Experimentally, the effect of quantum interference can be detected
by measuring the variation of the conductivity in an external magnetic
field. When the magnetic field is along the z-direction, $q_{x}$
and $q_{y}$ are quantized into a series of Landau levels as $q_{x}^{2}+q_{y}^{2}\to(n+\frac{1}{2})\ell_{B}^{-2}$
with $\ell_{B}=\sqrt{\frac{\hbar}{4eB}}$ the magnetic length and
$n$ a non-negative integer. Consequently, the magnetoconductivity
reads \cite{Note-on-SM} 
\begin{equation}
\delta\sigma_{\mathrm{qi}}(B)=\sum_{i=s,t_{\pm}}w_{i}\mathcal{F}\left(\frac{\ell_{B}^{2}}{\ell_{\phi}^{2}}+\frac{\ell_{B}^{2}}{\ell_{i}^{2}}\right)\label{eq: MC_single band}
\end{equation}
where $\mathcal{F}(x)\equiv\frac{e^{2}}{\pi h}[\psi(x+\frac{1}{2})-\ln x]$
with $\psi(x)$ the digamma function. Comparing with the previous
theories, only one Cooperon channel was taken into account in the
Hikami-Lukin-Nagaoka formula \cite{hikami1980spin} which is valid
only in two limits ($\eta=0$ and $\eta=1$). In Lu-Shen formula,
the two Cooperon channels of triple pairing $i=t_{\pm}$ were approximately
treated as one for WL, which forms a competition against the Cooperon
channel of singlet pairing ($i=s$) for WAL \cite{Lu_finite_T_prl_2014}.

When $\eta\ll1$, Eq. (\ref{eq: MC_single band}) is simplified as
$\delta\sigma_{\mathrm{qi}}(B)\approx-\frac{1}{2}\mathcal{F}(\frac{\ell_{B}^{2}}{\ell_{\phi s}^{2}})$
with an effective coherence length $\ell_{\phi s}$: $\frac{1}{\ell_{\phi s}^{2}}\simeq\frac{\eta^{2}}{\ell_{e}^{2}}+\frac{1}{\ell_{\phi}^{2}}$.
The presence of $\frac{\eta^{2}}{\ell_{e}^{2}}$ means a new decoherence
mechanism for the coherence length besides the interaction effect.
It is closely related to the correction to the $\pi$ Berry phase,
and becomes dominant at lower temperature as $\frac{1}{\ell_{\phi}^{2}}\rightarrow0$.
When $\eta$ is independent of the temperature, as shown in Fig. \ref{fig:zero-field-conductivity-correcti}(c),
the $\delta\sigma_{\mathrm{qi}}(B)$ gradually saturates when $\ell_{e}/\ell_{\phi}\to0$
as the effective coherence length is approximately determined by $\ell_{\phi s}=\ell_{e}/\eta$
instead of $\ell_{\phi}$ at low temperature. Hence, even a small
$\eta$ can generate an observable effect. When $0<1-\eta\ll1$, Eq.
(\ref{eq: MC_single band}) is simplified as $\delta\sigma_{\mathrm{qi}}(B)\approx\frac{1}{2}\mathcal{F}(\frac{\ell_{B}^{2}}{\ell_{\phi t_{+}}^{2}})$
with $\frac{1}{\ell_{\phi t_{+}}^{2}}=\frac{(1-\eta)^{2}}{4\ell_{e}^{2}}+\frac{1}{\ell_{\phi}^{2}}$,
where the new decoherence term $\frac{(1-\eta)^{2}}{4\ell_{e}^{2}}$
leads to the saturation of $\delta\sigma_{\mathrm{qi}}(B)$ when $\ell_{e}/\ell_{\phi}\to0$
{[}See Fig. \ref{fig:zero-field-conductivity-correcti}(d){]}.

This decoherence mechanism corresponds to the decaying Berry phase
of multiple scattering trajectories. The Berry phase contributes to
the return probability as a phase factor $e^{i\theta}=e^{i\phi_{b}(1+2n)}$
after $n$ times of revolutions \citep{gornyi2014interference}. For
$\eta\ll1$, after averaging over $n$ , we have $\langle e^{i\theta}\rangle\sim-e^{-\eta^{2}t/\tau}$,
where the minus sign stems from the $\pi$ Berry phase ($e^{i\pi(1+2n)}=-1$)
and gives a WAL correction when $\phi_{b}\sim\pi$. The decaying factor
can reproduce the effective coherence length $\ell_{\phi s}$ in the
magnetoconductivity formula for WAL \citep{Note-on-SM}. Furthermore,
in the magnetic TIs, $\eta$ can be a function of the temperature.
$\ell_{\phi s}$ or $\ell_{\phi t_{+}}$ can be a non-monotonic function
of temperature and further leads to a non-monotonic temperature dependence
of magnetoconductivity. In addition, $\delta\sigma_{\mathrm{qi}}(B)$
is still a monotonic function of the magnetic field. Thus, a temperature-dependent
$\eta$ can produce different temperature and magnetic field dependence
of magnetoconductivity.

\paragraph*{Fitting the experiment.}

Armed with the formula of magnetoconductivity in Eq. (\ref{eq: MC_single band}),
we are now ready to address the puzzle of the anomalous temperature
dependence of the conductivity. In Fig. \ref{fig:anomalous conductivity},
the experimental data labeled by open squares are extracted from the
temperature-dependent conductivity at finite $B$-field in Fig. 4(a)-(c)
in Ref. \citep{tkavc2019influence}. Since the conductivity correction
from the interaction effect is insensitive to the external magnetic
field, the magnetoconductivity $\delta\sigma(B)$ can exclude the
correction from the interaction effect and is mainly determined by
the quantum interference effect, $\delta\sigma(B)\approx\delta\sigma_{\mathrm{qi}}(B)$.
For the pristine $\mathrm{Bi_{2}Se_{3}}$ of $x_{\mathrm{Mn}}=0\%$,
the Fermi level insects with both the surface band and bulk bands
as clearly shown in the ARPES data in Ref. \citep{tkavc2019influence},
the $\delta\sigma$ data at different magnetic field can be well fitted
by considering one gapless surface states and two gapped bulk sub-bands(solid
red lines in Fig. \ref{fig:anomalous conductivity}a) \citep{Lu2011bulk,Garate2012bulksurface,Velkov2018bulksurface},
and the fitting details can be found in Ref. \citep{Note-on-SM}.

The magnetoconductivities of the samples of $x_{\mathrm{Mn}}=4\%$
and $x_{\mathrm{Mn}}=8\%$ are similar, and turn to increase with
decreasing temperature at low temperatures. The anomalous Hall resistivity
in a ferromagnetic conductor has an empirical relation with the magnetic
field $B$ and magnetization $M$, $\rho_{xy}=R_{0}B+R_{A}M$ \citep{Nagaosa2010AHE}.
The magnetization is a function of temperature below the Curie temperature
$T_{C}$. Nonzero magnetization makes the surface states open a tiny
gap. For the sample of $x_{\mathrm{Mn}}=8\%$, from the data of the
anomalous Hall resistivity, it is found that $M$ is proportional
to $1-\sqrt{\frac{T}{T_{C}}}$ below the Curie temperature $T_{C}=11.45\,\mathrm{K}$
\cite{Note1}. $\eta$ is assumed to obey the same behavior: $\eta(T)=\eta_{0}[1-\sqrt{\frac{T}{T_{C}}}]\Theta(T_{C}-T)$
(see Sec. SIII.B in Ref. \citep{Note-on-SM}), where $\eta_{0}$ is
the spin polarization at the zero temperature, and $\Theta(x)$ is
the Heaviside step function. Besides, the mean free path is estimated
as $\ell_{e}\approx14\,\mathrm{nm}$ at $T=2\,\mathrm{K}$ and $\ell_{e}\approx13.6\,\mathrm{nm}$
at $T=40\,\mathrm{K}$ from the mobility and carrier density data.
$\ell_{e}$ is insensitive to temperature and is fixed as $14\,\mathrm{nm}$
to reduce the number of fitting parameters. We further assume $\ell_{\phi}=\ell_{\phi}^{0}T{}^{-\delta/2}$
where $\ell_{\phi}^{0}$ and $\delta$ are the fitting parameters
and $T$ in unit of Kelvin. In Fig. \ref{fig:anomalous conductivity}(c),
the fitting curves show a good agreement with the experimental data
for $B=0.1,0.2,0.5$ and $1\,\mathrm{T}$. The corresponding fitting
parameters are listed in the Table SII in Ref. \citep{Note-on-SM}.
As the fitting parameter $\eta\simeq0.2$, the weighting factors $w_{i=t_{\pm}}\simeq4\eta^{2}$
and $w_{i=s}\simeq-\frac{1}{2}(1-8\eta^{2})$. Thus the Cooperon channel
of $i=s$ is dominant. Its effective phase coherence length $l_{\phi,i=s}$
has a non-monotonic temperature dependence, which is similar to the
one given by Tkac et al. \cite{tkavc2019influence}. A similar analysis
has been applied to the sample of $x_{Mn}=4\%$ in Ref. \cite{Note-on-SM},
and the fitting curves show a good agreement with the experimental
data for $B=0.2,0.5,1$ and $2\,\mathrm{T}$, as displayed in Fig.
\ref{fig:anomalous conductivity}(b). The good coincidence between
the theory and the experiment implies that the anomalous temperature
dependence of $\delta\sigma$ in the magnetic TIs can be ascribed
to the temperature-dependent $\eta$ or the Berry phase.

\begin{figure}
\begin{centering}
\includegraphics[clip,width=8cm]{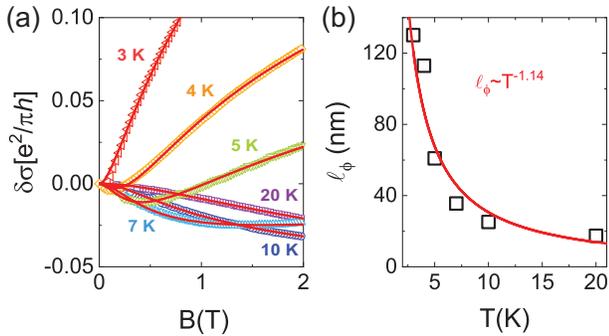} 
\par\end{centering}
\caption{\label{fig:Cr-doped TI} (a) Magnetoconductivity at different temperatures
for the Cr-doped $\mathrm{Bi_{2}Se_{3}}$ thin film of $x=0.23$.
The open squares are the experimental data extracted from Fig. 2(j)
of Ref. \cite{Liu_crossover_prl_2011}. The solid red lines are the
fitting results. (b) The temperature dependence of the fitted phase
coherence length $\ell_{\phi}$ (open squares). The red line indicates
$\ell_{\phi}\propto T^{-1.14}$.}
\end{figure}

Furthermore, we also applied the formula in Eq. (\ref{eq: MC_single band})
to fit the temperature-dependent magnetoconductivity in Cr-doped $\mathrm{Bi_{2}Se_{3}}$
ultrathin films in an early measurement \cite{Liu_crossover_prl_2011}
by considering two topological surface states and two gapped bulk
sub-bands. It was found that the measured crossover from WL to WAL
by increasing temperature can be well understood by taking into account
the temperature dependence of magnetization of the topological surface
states and quantum interference effect of multiple Cooperon channels.
Fig. \ref{fig:Cr-doped TI}(a) shows an excellent agreement between
the experimental data and fitting curves, and the corresponding fitting
parameters are also consistent at different temperatures. The extracted
phase coherence length follows the power law $\ell_{\phi}\propto T^{-1.14}$
{[}see Fig. \ref{fig:Cr-doped TI}(b){]}. More details are referred
to Sec. SIV in Ref. \citep{Note-on-SM}.

We would like to thank J. Honolka and K. Vyborny for providing original
experimental data in Fig. \ref{fig:anomalous conductivity}. This
work was supported by the Research Grants Council, University Grants
Committee, Hong Kong under Grant No. 17301717 and C7036-17W.

\bibliographystyle{apsrev} 

%%%%%%%%%%%%%%%%%%%%%%%%%%

\end{document}